\documentclass[conference]{IEEEtran}
\IEEEoverridecommandlockouts
\def\BibTeX{{\rm B\kern-.05em{\sc i\kern-.025em b}\kern-.08em
    T\kern-.1667em\lower.7ex\hbox{E}\kern-.125emX}}

\usepackage{enumerate}
\usepackage{enumitem}

\usepackage[utf8]{inputenc} 
\usepackage[T1]{fontenc}
\usepackage{hyperref}         
\usepackage{url}              
\usepackage{ifthen}           
\usepackage{cite}             

\usepackage[cmex10]{amsmath}  
\usepackage{amssymb}
\usepackage{braket}

\newboolean{JOURNAL} 
\setboolean{JOURNAL}{false} 

\usepackage{algorithm}
\usepackage{algpseudocode}

\algnewcommand{\IfThenElse}[3]{
  \State \algorithmicif\ #1\ \algorithmicthen\ #2\ \algorithmicelse\ #3}
  
\usepackage{comment}
\usepackage{bm}

\usepackage{tikz}	
\usepackage{xcolor}
\usetikzlibrary{backgrounds,fit,decorations.pathreplacing}  
\usetikzlibrary{automata} 
\usetikzlibrary{circuits.logic.CDH}
\usetikzlibrary{circuits.logic.US}
\usetikzlibrary{mindmap}
 
\usetikzlibrary{trees}

 \usetikzlibrary{shapes,decorations.pathreplacing}
 
\usetikzlibrary{arrows.meta}
\usetikzlibrary{shapes.geometric}

 \usepackage{xspace}

\usepackage{ulem}
\usepackage{enumitem}
\usepackage{mleftright}       
\mleftright                   

\usepackage{graphicx}         
\usepackage[space]{grffile}			
\usepackage{booktabs}         

\let\emph\textit
\usepackage{amsthm}
\theoremstyle{definition}		

\newtheorem{definitionenv}{Definition}
\newtheorem{lemmaenv}[definitionenv]{Lemma}
\newtheorem{theoremenv}[definitionenv]{Theorem}
\newtheorem{corollaryenv}[definitionenv]{Corollary}
\newtheorem{propositionenv}[definitionenv]{Proposition}
\newtheorem{conjectureenv}[definitionenv]{Conjecture}
\newtheorem{remarkenv}[definitionenv]{Remark}
\newenvironment{remark}{\begin{remarkenv}\rm}{\end{remarkenv}}
\newcommand{\br}{\begin{remark}}
	\newcommand{\er}{\end{remark}}

\newtheorem{exampleenv}{Example}
\newtheorem{app-lemmaenv}[section]{Lemma}

\newenvironment{definition}{\begin{definitionenv}\rm}{\end{definitionenv}}
\newenvironment{lemma}{\begin{lemmaenv}\rm}{\end{lemmaenv}}
\newenvironment{theorem}{\begin{theoremenv}\rm}{\end{theoremenv}}
\newenvironment{corollary}{\begin{corollaryenv}\rm}{\end{corollaryenv}}
\newenvironment{example}{\begin{exampleenv}\rm}{\end{exampleenv}}
\newenvironment{proposition}{\begin{propositionenv}\rm}{\end{propositionenv}}
\newenvironment{conjecture}{\begin{conjectureenv}\rm}{\end{conjectureenv}}
\newenvironment{app-lemma}{\begin{app-lemmaenv}\rm}{\end{app-lemmaenv}}

\newcommand{\bd}{\begin{definition}}
	\newcommand{\ed}{\end{definition}}
\newcommand{\bl}{\begin{lemma}}
	\newcommand{\el}{\end{lemma}}
\newcommand{\elp}{\hspace*{\fill} $\Box$
\end{lemma}}
\newcommand{\bt}{\begin{theorem}}
\newcommand{\et}{\end{theorem}}
\newcommand{\etp}{\hspace*{\fill} $\Box$
\end{theorem}}
\newcommand{\bc}{\begin{corollary}}
\newcommand{\ec}{\end{corollary}}
\newcommand{\ecp}{\hspace*{\fill} $\Box$
\end{corollary}}
\newcommand{\bcj}{\begin{conjecture}}
\newcommand{\ecj}{\end{conjecture}}

\newcommand{\be}{\begin{example}}
\newcommand{\ee}{\end{example}}
\newcommand{\eep}{\hspace*{\fill} $\Box$
\end{example}}
\newcommand{\bp}{\begin{proposition}}
\newcommand{\ep}{\end{proposition}}
\newcommand{\epp}{
\end{proposition}}

\newcommand{\cG}{{\cal G}}

\newcommand{\cS}{{\cal S}}

\usepackage{mathrsfs}
\newcommand{\code}[1]{\mathscr{#1}}
\newcommand{\group}[1]{\mathcal{#1}}

\newcommand{\st}[1]{\mathcal{#1}}

\usepackage{nicefrac}   
\newcommand{\Field}{\mathbb{F}}      
\newcommand{\eqdef}{\triangleq} 

\usepackage[colorinlistoftodos,prependcaption]{todonotes}

\usepackage{xcolor}
\usepackage{xargs}

\newcommandx{\yellownote}[2][1=]{\todo[inline,linecolor=yellow,backgroundcolor=yellow!25,bordercolor=yellow,#1]{#2}}
\newcommandx{\greennote}[2][1=]{\todo[inline,linecolor=green,backgroundcolor=green!25,bordercolor=green,#1]{#2}}

\begin{document}

\title{Advancing Finite-Length Quantum Error Correction Using Generalized Bicycle Codes\\
}

\author{
  \IEEEauthorblockN{Olai \AA.~Mostad\IEEEauthorrefmark{1}, Hsuan-Yin Lin\IEEEauthorrefmark{1}, Eirik~Rosnes\IEEEauthorrefmark{1}, De-Shih Lee\IEEEauthorrefmark{2},  and Ching-Yi Lai\IEEEauthorrefmark{2}}
  \IEEEauthorblockA{\IEEEauthorrefmark{1}\textit{Simula UiB}, N--5006 Bergen, Norway, \{olai, lin, eirikrosnes\}{@}simula.no}
  \IEEEauthorblockA{\IEEEauthorrefmark{2}%
    \textit{Institute of Communications Engineering, National Yang Ming Chiao Tung University}, \\ Hsinchu 300093, Taiwan, \{sky620529{@}gmail.com, cylai{@}nycu.edu.tw\}}
}

\maketitle

\begin{abstract}
 
Generalized bicycle (GB) codes have emerged as a promising class of quantum error-correcting codes with practical decoding capabilities. While numerous asymptotically good quantum codes and quantum low-density parity-check code constructions have been proposed, their finite block-length performance often remains unquantified. In this work, we demonstrate that GB codes exhibit comparable or superior error correction performance in finite-length settings, particularly when designed with higher or unrestricted row weights. Leveraging their flexible construction, GB codes can be tailored to achieve high rates while maintaining efficient decoding. We evaluate GB codes against other leading quantum code families, such as quantum Tanner codes and single-parity-check product codes, highlighting their versatility in practical finite-length applications.


\end{abstract}


\section{Introduction}

Quantum error correction (QEC) plays a vital role in protecting quantum information from noise, enabling fault-tolerant quantum computation and  network communication~\cite{Terhal15_1, LaiKuo24_1}. QEC codes achieve this by encoding a few logical qubits into multiple physical qubits. Among these, topological codes have demonstrated comprehensive procedures for practical applications~\cite{Kitaev03_1, DennisKitaevLandahlPreskill02_1, FowlerMariantoniMartinisCleland12_1}. Recent advancements in quantum low-density parity-check (LDPC) codes~\cite{Gottesman14_1} have introduced more efficient methods for achieving reliable quantum computation~\cite{Bravyi-etal24_1, LinPryadko24_1}. Notably, asymptotically good quantum LDPC codes have been proven to exist~\cite{PanteleevKalachev22_1, PanteleevKalachev22_2,TillichZemor14_1}, and quantum Tanner (QT) codes~\cite{LeverrierZemor22_1, LeverrierZemor23_1, MostadRosnesLin24_1} offer explicit constructions of such codes within the LDPC framework.

Among the diverse families of QEC codes, the search for practical constructions that balance error correction performance, finite-length feasibility, and efficient decoding is ongoing. This work aims to advance the potential of generalized bicycle (GB) codes~\cite{MacKayMitchisonMcFadden04_1, KovalevPryadko13_1, PanteleevKalachev21_1}, a class of quantum codes constructed using pairs of circulant matrices. A straightforward procedure for constructing even-length GB codes with specified parameters is introduced. The flexibility of GB codes supports the design of high-rate codes, making them particularly appealing for finite-length regimes. Unlike many asymptotically good quantum code constructions with limited parameter flexibility, GB codes offer a more practical and versatile solution.

We demonstrate the potential of GB codes by constructing several finite-length examples and comparing their performance against other leading quantum code families, such as QT codes, bivariate bicycle (BB) codes~\cite{Bravyi-etal24_1}, and single-parity-check (SPC) product codes~\cite{OstrevOrsucciLazaroMatuz24_1}. Our findings reveal that GB codes offer competitive or superior QEC  performance when restricted to similar row weights in the parity-check matrix. Moreover, without row-weight constraints, GB codes can achieve exceptional performance in both high-error-rate and low-error-rate regimes. This is particularly evident when increased or unrestricted row weights are leveraged to enhance the minimum distance.


 


The decoding performance of finite-length quantum LDPC codes is evaluated using the  memory belief propagation (MBP) decoding algorithm~\cite{KuoLai22_2} combined with postprocessing via approximate degenerate ordered statistics decoding (ADOSD)~\cite{KungKuoLai24_1sub}.
By employing this general decoding strategy for quantum LDPC codes, this work provides a systematic and unbiased analysis of their performance.

The remainder of the paper is organized as follows. Section II provides an overview of quantum stabilizer codes. Section III explores various code constructions, including GB codes, QT codes, SPC product codes, and BB codes. Section IV presents new GB codes along with detailed simulation results, comparing their performance against other quantum code families. Finally, Section V concludes the paper.

\section{Quantum Stabilizer Codes}
 
Let $I$, $X$, $Y$, and $Z$ denote the single-qubit Pauli matrices. The $n$-fold Pauli group $\mathcal{G}_n$ consists of all $n$-fold tensor products of Pauli matrices with an overall phase factor in $\{\pm 1, \pm \sqrt{-1}\}$, acting naturally on the Hilbert space $\mathbb{C}^{2^n}$, the $2^n$-fold Cartesian product of the complex field. The weight of a Pauli operator $E \in \mathcal{G}_n$ 
is defined as the number of tensor factors in $E$ that are not equal to the identity operator $I$.

In this paper, we consider the code capacity model, assuming independent depolarizing errors with rate $\epsilon$, where each qubit independently suffers an $X$, $Y$, or $Z$ error with probability $\nicefrac{\epsilon}{3}$ and no error with probability $1-\epsilon$.
  
A stabilizer code $\code{C}(\mathcal{S}) \subseteq \mathbb{C}^{2^n}$ is the joint $+1$ eigenspace of an Abelian subgroup $\mathcal{S} \subset \mathcal{G}_n$ with $-I^{\otimes n} \notin \mathcal{S}$~\cite{NielsenChung10_1}, where $\otimes$ denotes the Kronecker product. If $\mathcal{S}$ has $n-k$ independent generators, then $\code{C}(\mathcal{S})$ is an $[[n,k,d]]$ code encoding $k$ logical qubits into $n$ physical qubits. The minimum distance $d$ is the minimum weight of a nontrivial Pauli operator that acts nontrivially on the code space.

An error $E\in\cG_n$ can be detected by $\code{C}(\cS)$ through stabilizer measurements if $E$ anticommutes with at least one of its stabilizers. Let $\{g_i\}_{i=1}^m$, where $m \geq n-k$, be a set of generators for $\mathcal{S}$. The binary measurement outcomes of these generators constitute the \emph{error syndrome} of $E$. Since the stabilizer group acts trivially on the code space, it introduces a form of code degeneracy. 
A stabilizer code is said to be \textit{degenerate} if it has stabilizers with weights smaller than its minimum distance. A quantum code is considered \emph{highly degenerate} when it contains numerous stabilizers with weights below its minimum distance.

We focus on Calderbank–Shor–Steane (CSS) codes~\cite{CalderbankShor96_1, Steane96_1}, which are defined by a stabilizer group consisting of stabilizers that are exclusively of either $X$-type or $Z$-type. We denote the binary field by $\Field_2\eqdef\{0,1\}$. An $[[n, k]]$ CSS code is represented by two separate binary parity-check matrices $H_{\textnormal{X}} \in \Field_2^{m_{\textnormal{X}} \times n}$ and $H_{\textnormal{Z}} \in \Field_2^{m_{\textnormal{Z}} \times n}$, where $m_{\textnormal{X}} + m_{\textnormal{Z}} \geq n - k$  and the matrices satisfy the orthogonality condition
\begin{equation}
  H_{\textnormal{X}} H_{\textnormal{Z}}^\top = 0, \label{eq:orthogonality}
\end{equation}
where the superscript ``${\scriptstyle\top}$" denotes the transpose operator. Each binary vector in the row space of $H_{\textnormal{X}}$ corresponds to an $X$-type stabilizer, with the entries indicating the nontrivial support of the stabilizer. For example, the binary vector $10010$ corresponds to the stabilizer $X \otimes I \otimes I \otimes X \otimes I$. Similarly, the binary vectors in the row space of $H_{\textnormal{Z}}$ define the $Z$-type stabilizers.
Thus, the dimension \( k \) of the CSS code is
\begin{equation*}
  k = n - \textrm{rank}(H_{\textnormal{X}}) - \textrm{rank}(H_{\textnormal{Z}}).
\end{equation*}
The minimum distance of CSS codes can be determined using the procedure proposed in
\cite{RosnesYtrehus09_1, RosnesYtrehusAmbrozeTomlinson12_1}.

\subsection{Decoders}



For decoding performance, we employ the quaternary MBP$_4$+ADOSD$_4$ decoder~\cite{KungKuoLai24_1sub}.
The postprocessing algorithm, ADOSD$_4$, leverages the hard-decision history of MBP$_4$ along with the output belief distributions of errors at each location to perform ordered statistics decoding (OSD). Additionally, it identifies certain highly reliable bits, which allow for a reduction in the size of the syndrome decoding problem, enhancing computational efficiency. Moreover, certain degenerate conditions under which high-order OSD reduces to zero-order OSD are identified.  For further details, see~\cite{KungKuoLai24_1sub}.

\section{Code Constructions}


\subsection{Generalized Bicycle Codes}
 GB codes \cite{KovalevPryadko13_1,PanteleevKalachev21_1} extend the concept of bicycle codes originally introduced by MacKay \textit{et al.} in \cite{MacKayMitchisonMcFadden04_1}. A bicycle code is constructed by first selecting $H^\prime=[M \quad M^\top ]$, where $M$ is a random square circulant matrix and then  defining $H_{\textnormal{X}}=H_{\textnormal{Z}}=H'$  
 after removing rows from  $H^\prime$ to achieve the desired dimensions.

 GB codes generalize this approach by allowing any two commuting matrices, 
 $M_1$ and $M_2$, to be chosen. A GB code is then defined as
\begin{equation}
    H_{\textnormal{X}}=[M_1\quad M_2],\quad H_{\textnormal{Z}}=[M_2^\top\quad M_1^\top]. \label{eq:GB_construction}
\end{equation}
This approach offers greater flexibility in designing quantum codes. In particular, $M_1$ and $M_2$ can be chosen as two circulant matrices, ensuring that the orthogonality condition~(\ref{eq:orthogonality}) is satisfied.
In this case, all rows of $H_{\textnormal{X}}$ and $H_{\textnormal{Z}}$ have a uniform row weight $\rho$.

 To construct an even-length $[[n=2\ell, k]]$ GB code with row weight $\rho$,   we consider the following two approaches. 

\subsubsection{Construction MMM}
\label{sec:construction-mmm}
In \cite{KovalevPryadko13_1,PanteleevKalachev21_1}, the dimension of a GB code with circulant matrices $M_1$ and $M_2$ is determined by analyzing the associated polynomials (see Construction PK below). To enable comparisons with codes of various parameters discussed in this work, we propose a straightforward method for constructing GB codes, following the bicycle code construction described in \cite{MacKayMitchisonMcFadden04_1}.
This procedure, referred to as \textit{Construction MMM}, is outlined below.
\begin{enumerate}[label=(\roman*)]
\item Generate two $\ell \times \ell$ random circulant matrices $M_1$ and $M_2$, with row weights $\rho_1$ and $\rho_2$, respectively,
such that  $ \rho_1 + \rho_2=\rho$.
 
\item Define $H'_{\textnormal{X}} = [M_1 \quad M_2]$ and $H'_{\textnormal{Z}} = [M_2^\top \quad M_1^\top]$, and obtain the full-rank matrices $H_{\textnormal{X}}$ and $H_{\textnormal{Z}}$ by removing redundant rows from $H'_{\textnormal{X}}$ and $H'_{\textnormal{Z}}$, respectively. 

\item Remove excess rows from $H_{\textnormal{X}}$ and $H_{\textnormal{Z}}$ to ensure that $\textrm{rank}(H_{\textnormal{X}}) + \textrm{rank}(H_{\textnormal{Z}}) = n - k$.

\end{enumerate}

It is important to maintain uniform column weights during the row removal process. The resulting GB code will have all rows of $H_{\textnormal{X}}$ and $H_{\textnormal{Z}}$ with a row weight $\rho = \rho_1 + \rho_2$.

This procedure enables the construction of GB codes with any even code length. Furthermore, the row weight is not constrained to be either odd or even.

\subsubsection{Construction PK}
\label{sec:construction-pk}

Alternatively, we use the approach proposed in \cite[Sec.~4]{PanteleevKalachev21_1} to construct low-rate GB codes, referred to as \emph{Construction PK}.

In this method, the circulant matrices $M_1$ and $M_2$ in (\ref{eq:GB_construction}) are derived from two polynomials, $a(x)$ and $b(x)$, respectively. The code dimension is determined as $k = 2\deg(g(x))$, where $g(x) \triangleq \gcd\bigl(a(x), b(x), x^\ell - 1\bigr)$, the greatest common divisor of the three polynomials~\cite[Prop.~1]{PanteleevKalachev21_1}. For our construction,   we first randomly generate a polynomial $g(x)$ such that its degree $\deg(g(x)) = \nicefrac{k}{2}$ and $g(x) \mid (x^\ell - 1)$, i.e., $g(x)$ is a divisor of $x^\ell - 1$. Once $g(x)$ is determined, we randomly generate polynomials $a(x)$ and $b(x)$ with row weights $\rho_1$ and $\rho_2$ such that $g(x) \mid a(x)$ and $g(x) \mid b(x)$.

It is important to note, as indicated in~\cite[Sec.~4.4]{PanteleevKalachev21_1}, that approximately $2^{\deg(g(x))}$ trials are required to find a polynomial divisible by $g(x)$. Consequently, $\deg(g(x))$ must remain small, leading to a small code dimension $k$, and therefore to low-rate GB codes.


 
\subsection{Quantum Tanner Codes}
QT codes \cite{LeverrierZemor22_1,LeverrierZemor23_1} represent a significant advancement in the development of asymptotically good families of quantum codes. These codes are constructed by combining two classical codes with the left-right Cayley complex introduced in \cite{DinurEvraLivneLubotzkyMozes22_1}, which extends the concept of expander codes \cite{SipserSpielman96_1} into the quantum domain.

The construction begins with a group \( \group{G} \) and  two generating sets, \( \st{A} = \st{A}^{-1} \subseteq \mathcal{G}\) and \( \st{B} = \st{B}^{-1} \subseteq \mathcal{G} \), where \( \vert \st{A} \vert = \vert \st{B} \vert = \Delta \), for some integer $\Delta>0$. We construct a complex composed of vertices $\st{V}$, \( \st{A} \)-edges, \( \st{B} \)-edges, and squares.

In the quadripartite version \cite{LeverrierZemor23_1}, the vertex set \( \st{V} \) is partitioned into four subsets as follows,
\begin{equation*}
  \st{V} = \st{V}_{00} \cup \st{V}_{01} \cup \st{V}_{10} \cup \st{V}_{11}, \quad \st{V}_{ij} = \group{G} \times \{ij\}, \ i,j \in \{0,1\}.
\end{equation*}
Also, define \(\st{V}_0 \triangleq \st{V}_{00} \cup \st{V}_{11} \) and \( \st{V}_1 \triangleq \st{V}_{01} \cup \st{V}_{10} \). The \( \st{A} \)-edges are \( \{((g,i0), (ag,i1))\colon g\in \group{G}, a\in \st{A}, i\in\{0, 1\}\} \), the \( \st{B} \)-edges are  \( \{((g,0j), (gb,1j))\colon g\in \group{G}, b\in \st{B}, j\in\{0, 1\}\} \), and the squares are
\(
  \{\left[(g,00), (ag,01), (gb,10), (agb,11)\right]\colon a\in \st{A}, g\in \group{G}, b\in \st{B}\}.    
\)
Note that the construction prevents squares from collapsing, so we avoid the requirement \( ag \neq gb \) in the quadripartite version.

Let \( H_\textnormal{A} \in \mathbb{F}_2^{m \times \Delta} \) and \( H_\textnormal{B} \in \mathbb{F}_2^{(\Delta-m) \times \Delta} \) represent full-rank parity-check matrices of classical codes \( \code{C}_\textnormal{A} \) and \( \code{C}_\textnormal{B} \), respectively. Similarly, \( H_\textnormal{A}^\perp \in \mathbb{F}_2^{(\Delta-m) \times \Delta} \) and \( H_\textnormal{B}^\perp \in \mathbb{F}_2^{m \times \Delta} \) are parity-check matrices of the dual codes \( \code{C}_\textnormal{A}^\perp \) and \( \code{C}_\textnormal{B}^\perp \), respectively. Qubits are placed on squares, so the code length \( n \) of the QT code equals the number of squares, \( \vert \group{G} \vert \Delta^2 \), since each vertex is incident to \( \Delta^2 \) squares. Constraints are placed on vertices: for \( v \in \st{V}_0 \), all incident squares (qubits) must form a codeword of \( (\code{C}_\textnormal{A} \otimes \code{C}_\textnormal{B})^\perp \), corresponding to parity-checks of \( H_{\textnormal{X}} \). Similarly, for \( v' \in \st{V}_1 \), all incident squares must form a codeword of \( (\code{C}_\textnormal{A}^\perp \otimes \code{C}_\textnormal{B}^\perp)^\perp \), corresponding to parity-checks of \( H_{\textnormal{Z}} \). As a result, \( H_{\textnormal{X}} \) and \( H_{\textnormal{Z}} \) each contain \( \Delta(\Delta-m)\vert \group{G} \vert \) rows. Determining the exact dimension of the code requires analyzing the rank of
\( H_{\textnormal{X}} \) and \( H_{\textnormal{Z}} \). The properties of QT codes strongly depend on the expansion quality of the associated Cayley graphs $\textnormal{Cay}_\textnormal{l}(\group{G}, \st{A})$, $\textnormal{Cay}_\textnormal{r}(\group{G}, \st{B})$ and the robustness of the dual tensor codes \( (\code{C}_\textnormal{A} \otimes \code{C}_\textnormal{B})^\perp \) and \( (\code{C}_\textnormal{A}^\perp \otimes \code{C}_\textnormal{B}^\perp)^\perp \) \cite{LeverrierZemor23_1}.

Note that the $\st{A}$-edge connecting the vertices $(g, i0)$ and $(ag, i1)$ is viewed as coming from $a\in \st{A}$ when placing restrictions on $(g, i0)$, but as  its inverse $a^{-1}\in \st{A}$ when placing restrictions on $(ag, i1)$. It turns out that it can also be viewed as coming from $a$ on both vertices~\cite{MostadRosnesLin25_1sub}, but for the codes we construct here, this seems to be detrimental to the minimum distance.

For the constructions considered, let \( \group{G} = \mathbb{Z}/8\mathbb{Z} \) be the additive group of integers modulo $8$, with generators \( \st{A} = \st{B} =\{g\in \group{G}: g \neq 0\} \). The code \( \code{C}_{\textnormal{A}} \) is the \([7,4,3]\) Hamming code, and \( \code{C}_{\textnormal{B}} = \code{C}_{\textnormal{A}}^\perp \). 
From two specific (orthogonal) parity-check matrices for the Hamming code and its dual, we obtained  \( [[392,32,12]] \) and \([[392,48,12]]\) QT codes with row weights $12$ and $16$, and an average row weight $\bar{\rho}$ of $13.03$ and $13.05$, respectively.
It was suggested in~\cite{LeverrierZemor22_1} that QT codes constructed in this way could exhibit strong performance. While the theoretical bound on the dimension is only $8$, we get codes with dimensions $32$ and $48$. Note that the ordering of the columns in the Hamming parity-check matrix significantly influences the construction, and we were unable to improve upon the $[[392,48,12]]$ parameters even when independently permuting the columns of the parity-check matrices, possibly resulting in $\code{C}_\textnormal{B}\neq \code{C}_\textnormal{A}^\perp$.

%

\subsection{Single-Parity-Check Product Codes}
Product CSS codes extend the concept of classical product codes to quantum error-correcting codes, offering a relatively high code rate and a simple structural design~\cite{OstrevOrsucciLazaroMatuz24_1}. A family of SPC $D$-fold product codes has been proposed, demonstrating strong performance compared to other codes with similar code rates.
For example, a $[[512,174,8]]$ $3$-fold SPC  product code can be  defined by the parity-check matrices 
\renewcommand{\arraycolsep}{1pt} 
\begin{align*} 
H_{\textnormal{X}} =& \left(\begin{array}{ccccccccccccccccc}
    h&\otimes &h&\otimes &h&\otimes &I&\otimes &I&\otimes &I&\otimes &I&\otimes &I&\otimes &I \\
    I&\otimes &I&\otimes &I&\otimes &h&\otimes &h&\otimes &h&\otimes &I&\otimes &I&\otimes &I \\
    I&\otimes &I&\otimes &I&\otimes &I&\otimes &I&\otimes &I&\otimes &h&\otimes &h&\otimes &h
\end{array}\right),\\
H_{\textnormal{Z}} =& \left(\begin{array}{ccccccccccccccccc}
    h&\otimes &I&\otimes &I&\otimes &h&\otimes &I&\otimes &I&\otimes &h&\otimes &I&\otimes &I \\
    I&\otimes &h&\otimes &I&\otimes &I&\otimes &h&\otimes &I&\otimes &I&\otimes &h&\otimes &I \\
    I&\otimes &I&\otimes &h&\otimes &I&\otimes &I&\otimes &h&\otimes &I&\otimes &I&\otimes &h
\end{array}\right),
\end{align*}   
where $h=(1\quad 1)$. 
Notably, all the rows of both $H_{\textnormal{X}}$ and $H_{\textnormal{Z}}$  have a uniform row weight of $8$.
 
\subsection{Bivariate Bicyle Codes}

BB codes are a type of GB code constructed using base matrices derived from bivariate polynomials \cite{Bravyi-etal24_1}. These codes are quantum LDPC codes that exhibit error thresholds comparable to surface codes, yet they require significantly fewer physical qubits.

Within the framework of GB codes, BB codes naturally conform to the construction in (\ref{eq:GB_construction}). Consider the following two matrices:
\begin{equation*}
  U = S_{\ell_1} \otimes I_{\ell_2}\quad\textnormal{and}\quad W = I_{\ell_1} \otimes S_{\ell_2},
\end{equation*}
where $I_{\ell_i}$ is the $\ell_i \times \ell_i$ identity matrix  and $S_{\ell_i}$ is a right cyclic shift of $I_{\ell_i}$. The base matrices $M_1$ and $M_2$ of the BB code in (\ref{eq:GB_construction}) are defined as
\begin{equation*}
  M_1 = M_{11} + M_{12} + M_{13}
  \quad\textnormal{and}\quad
  M_2 = M_{21} + M_{22} + M_{23},
\end{equation*}
where each $M_{ij}$ is a power of either $U$ or $W$. Since $U$ and $W$ are circulants, with $U^{\ell_1} = W^{\ell_2} = I_{\ell_1\ell_2}$ and $UW = WU$, it follows that $M_{ij} \neq M_{ij^\prime}$ for $j \neq j^\prime$. Consequently, the parity-check matrices of the BB code have all rows of weight~$6$.


\section{New GB Codes and Performance Simulations}
We use the MBP$_4$+ADOSD$_4$ decoder~\cite{KungKuoLai24_1sub} with a parallel schedule and a maximum of $100$ BP iterations in all simulations. The parameter $\alpha$ in MBP is optimized by sweeping values from $1.0,1.1,\ldots$, until performance saturates in the low-error-rate regime. This tuning is code-dependent and can be done via pre-simulations,
independent of specific error instances. For each data point, at least $200$ error events are collected, or $100$ when the logical error rate falls below $10^{-4}$. For each code of minimum distance $2t+1$ or $2t+2$, we include a reference curve $y = ax^{t+1}$.
To benchmark CSS codes, we construct GB codes of the same length and dimension. Each GB code is generated randomly from   $5$–$20$ attempts, and the best performing code is selected for evaluation.
All simulated codes use full-rank parity-check matrices. Although this ensures a clean comparison, adding redundant checks (e.g.,  metachecks or overcomplete checks~\cite{KuoLai20_1, PanteleevKalachev21_1, MiaoSchnerringLiSchmalen25_1}) may further improve performance and is a direction for future work.


\subsection{Quantum Tanner Codes}



To compare the performance of the $[[392,32,12]]$ and $[[392,48,12]]$ QT codes, both with row weights $12$ and $16$, and an average row weight $\bar{\rho} \approx 13$, 
we construct a $[[392,32,14]]$ GB\#$1$ code with row weight $\rho=12$, and a $[[392,48,13]]$ GB\#$2$ code with row weight $\rho=13$, using the following polynomials based on Constructions PK and MMM, respectively,
\begin{IEEEeqnarray*}{rCl}
  \textnormal{GB}\#1 (\textrm{PK})&&
  \begin{cases}
    &\hspace{-2mm}1+x^{82}+x^{96}+x^{112}+x^{114}+x^{156},
    \\
    &\hspace{-2mm}1+x^{17}+x^{76}+x^{101}+x^{118}+x^{126}, 
  \end{cases}
  \\[1mm]
  \textnormal{GB}\#2 (\textrm{MMM})&&
  \begin{cases}
    &\hspace{-2mm}1+x^{27}+x^{29}+x^{50}+x^{107}+x^{161},
    \\
    &\hspace{-2mm}1+x^{38}+x^{48}+x^{50}+x^{67}+x^{158}+x^{179}. 
  \end{cases}\IEEEeqnarraynumspace
\end{IEEEeqnarray*}



\begin{figure}[t!]
  \centering
  \includegraphics[width=0.95\columnwidth, height=7.00cm]{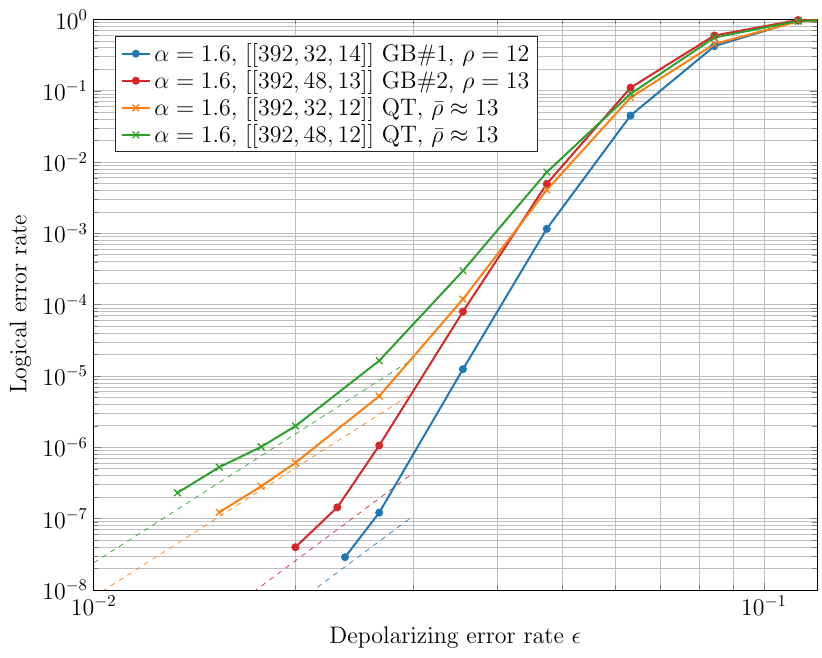}
  \caption{Decoding performance of the $[[392,32,12]]$ and $[[392,48,12]]$ QT codes, along with the $[[392,32,14]]$ GB\#$1$ and $[[392,48,13]]$ GB\#$2$ codes.}
  \label{fig:QT392}
\end{figure}

The comparison of these three codes of length $392$ is shown in Fig.~\ref{fig:QT392}.
It is evident that the GB\#$1$ code, with constant row weight $\rho=12$, significantly outperforms the $[[392,32,12]]$ QT code due to its higher minimum distance.
Similarly, the GB\#$2$ code, with constant row weight $\rho=13$, outperforms the $[[392,48,12]]$ QT code, with average row weight $\bar{\rho} \approx 13$. More importantly, no error floor is observed for these two GB codes, even at a simulated logical error rate of  $10^{-7}$. As expected, the $[[392,32,12]]$ QT code performs better than the $[[392,48,12]]$ QT code for all error rates.

\subsection{Single-Parity-Check Codes}
 
In~\cite{OstrevOrsucciLazaroMatuz24_1}, a $[[512,174,8]]$ SPC product code is shown to outperform certain bicycle codes, QT codes, and others with comparable parameters. Under the constraint of row weight $\rho=8$, we can construct a $[[512,174,6]]$ GB\#$3$ code  via Construction MMM using the  polynomials
\begin{equation*} 
\textnormal{GB}\#3 (\textrm{MMM})
\begin{cases}
    &\hspace{-2mm}1+x^{73}+x^{192}+x^{234},\\
    &\hspace{-2mm}1+x^{97}+x^{144}+x^{193}. 
\end{cases}
\end{equation*}
 
If the row weight constraint is relaxed to $\rho=9$, we can construct a $[[512,174,7]]$ GB\#$4$ code via Construction MMM using the  polynomials 
\begin{equation*} 
  \textnormal{GB}\#4 (\textrm{MMM})
  \begin{cases}
      &\hspace{-2mm} 1+x^{89}+x^{159}+x^{218},\\
      &\hspace{-2mm} 1+x^{20}+x^{61}+x^{103}+x^{131}.
  \end{cases}
\end{equation*}

\begin{figure}[t!]
    \centering
    \includegraphics[width=0.95\columnwidth, height=7.00cm]{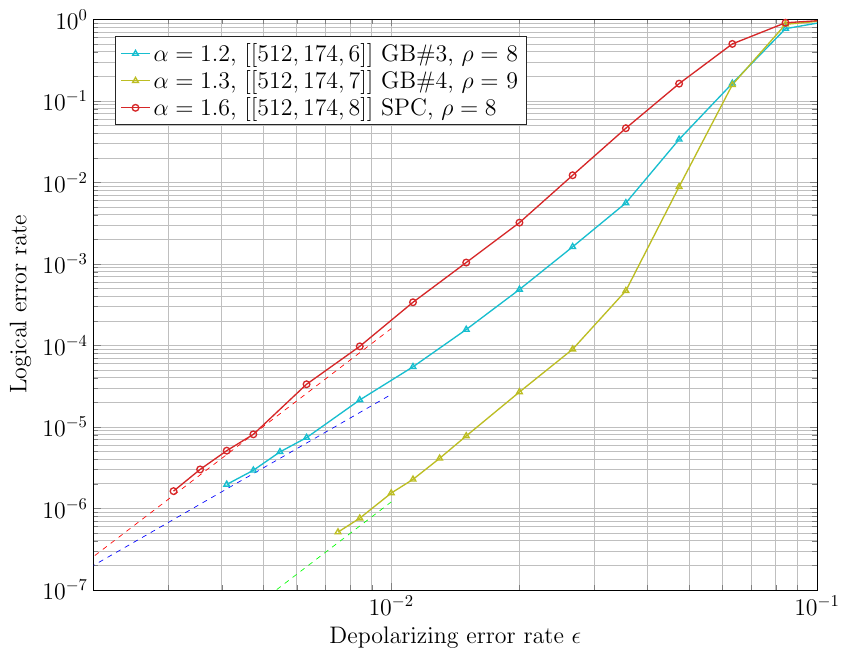}
    \vspace*{-0.50mm}
    \caption{Decoding performance of the $[[512,174,8]]$ SPC product code, along with  the $[[512,174,6]]$ GB\#$3$ and $[[512,174,7]]$ GB\#$4$ codes.}
    \label{fig:SPC512}
\end{figure}

Fig.~\ref{fig:SPC512} compares these three $[[512,174]]$ codes. Surprisingly, both GB codes outperform the $[[512,174,8]]$ SPC product code for depolarizing rates larger than $0.002$ in our simulations. With a slightly higher row weight of $\rho=9$, the GB\#$4$ code outperforms the SPC product code by about two orders of magnitude. Even the GB\#$3$ code, which has a smaller minimum distance of $6$, performs better, although the SPC product code is expected to outperform GB\#$3$ at depolarizing rates below $10^{-3}$ because of its higher minimum distance.

\subsection{Bivariate Bicycle Codes}

Finally, we compare the performance of the $[[144,12,12]]$ BB code from~\cite{Bravyi-etal24_1} with that of a GB code. We were only able to construct $[[144,12]]$ GB codes with minimum distances up to $8$ using row weights $\rho = 6$ or $\rho = 7$. However, by increasing the row weight to $\rho = 8$, we succeeded in constructing a $[[144,12,12]]$ GB code, referred to as GB\#$5$, as follows, 
\begin{equation*} 
\textrm{GB}\#5 (\textrm{PK}) 
\begin{cases}
    &\hspace{-2mm} 1+x^{3}+x^{32}+x^{47},\\
    &\hspace{-2mm} 1+x^{20}+x^{59}+x^{63}. 
\end{cases}
\end{equation*}

\begin{figure}[t!]
  \centering
  \includegraphics[width=0.95\columnwidth, height=6.75cm]{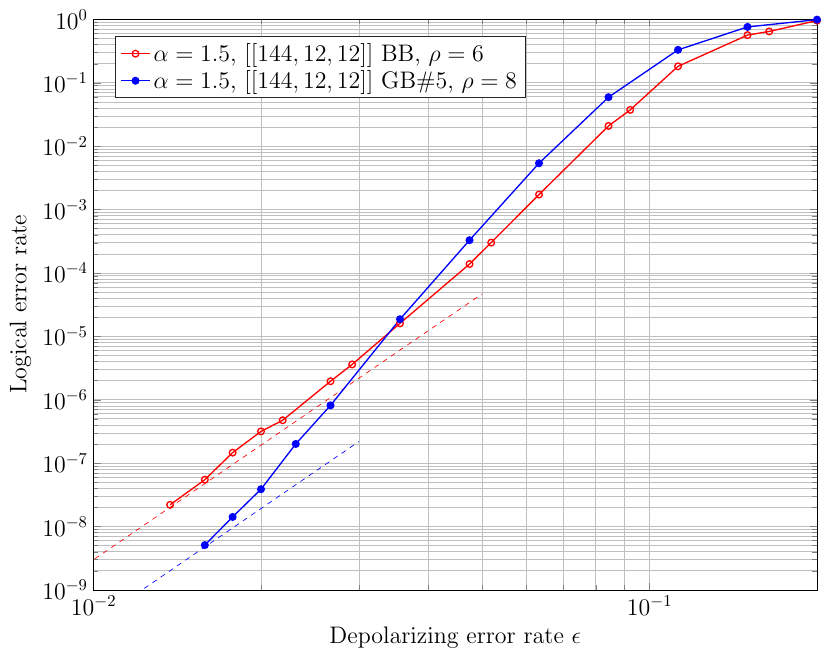}
  \vspace*{-2.00mm}
  \caption{Decoding performance of the $[[144,12,12]]$ BB code and the $[[144,12,12]]$ GB\#$5$ code.}
  \label{fig:BB144}
\end{figure}

Fig.~\ref{fig:BB144} presents a comparison of  the $[[144,12,12]]$ BB code and the $[[144,12,12]]$ GB\#$5$ code. 
 The results show that the BB code performs better at higher physical error rates, whereas the GB\#$5$ code outperforms the BB code at physical error rates below $0.03$. 
Since both codes have the same minimum distance, this suggests that the GB\#$5$ code
has a lower error floor, which is a highly desirable feature.


 
\section{Conclusion}
\label{sec:conclusion}
In this work, we explored the potential of  GB codes to advance finite-length quantum error correction. Through  Constructions MMM and PK, we demonstrated the flexibility of GB codes in achieving various code parameters with tailored row weights and minimum distances. Our simulations highlighted the competitive decoding performance of GB codes, which in many cases surpassed that of established families like QT codes and SPC product codes at comparable lengths and dimensions. 
While challenges remain in constructing high minimum distance GB codes for specific parameters, this work showcases the efficacy and adaptability of GB codes of practical size.

On the other hand, certain GB codes require stabilizer measurements with slightly higher row weights than the compared CSS codes. 
In particular, bare syndrome extraction would compromise its error correction performance under circuit-level noise.  Consequently, to fully realize the potential of the GB\#$5$ code in practical implementations, more sophisticated fault-tolerant techniques, such as flagged syndrome extraction~\cite{ChaoReichardt18_1, LiouLai24_1sub}, are necessary.
These results underscore an important trade-off between performance and implementation complexity in quantum code design.



\bibliographystyle{IEEEtran}
\bibliography{defshort1.bib, biblioHY.bib}

\end{document}